# An Enhanced Corpus for Arabic Newspapers Comments

Hichem Rahab[1], Abdelhafid Zitouni[2] and Mahieddine Djoudi[3]
[1]ICOSI Laboratory, University of Khenchela, Algeria
[2]LIRE Laboratory, University of Constantine 2, Algeria
[3]TechNE Laboratory, University of Poitiers, France

**Abstract:** *In this paper, we propose our enhanced approach to create a dedicated corpus for Algerian Arabic newspapers comments. The developed approach has to enhance an existing approach by the enrichment of the available corpus and the inclusion of the annotation step by following the Model Annotate Train Test Evaluate Revise (MATTER) approach. A corpus is created by collecting comments from web sites of three well know Algerian newspapers. Three classifiers, support vector machines, naïve Bayes, and k-nearest neighbors, were used for classification of comments into positive and negative classes. To identify the influence of the stemming in the obtained results, the classification was tested with and without stemming. Obtained results show that stemming does not enhance considerably the classification due to the nature of Algerian comments tied to Algerian Arabic Dialect. The promising results constitute a motivation for us to improve our approach especially in dealing with non-Arabic sentences, especially Dialectal and French ones.*

**Keywords:** *Opinion mining, sentiment analysis, K-Nearest Neighbours, Naïve Bayes, Support Vector Machines, Arabic, comment.*

---

## 1. Introduction

With the development of the web and its offered services, a huge amount of data is generated [25] and additional needs emerge to take benefit from this information thesaurus. Mining the political, economic and social opinion, to make the available information amount in an easily understood form by decision makers in dedicated centers is an emergent need in this air. Sentiment analysis vocation is to classify people opinions into specific classes to facilitate understanding the behind phenomenon.

A variety of classifications are available, some works deal only with positive vs. negatives classes [6, 32], other works deal with more important number of classes [12].

A very important amount of useful information is available in term of comments of newspapers websites visitors around the world and in different languages. A lot of works is this era deal with English language, and other European languages, but works treating Arabic language still in their beginning [5].

Arabic is a Semitic language spoken by about 300 million of people in 22 Arab countries. The importance of Arabic is also that it is the language of the holy Quran [12] the book of 1.5 billion Muslim in the world. We can find three forms of Arabic language, Classical Arabic, Modern Standard Arabic (MSA), and Dialectal Arabic. Classical Arabic is the original form of the language preserved from centuries by the Islamic literature and especially the holy Quran. For Modern Standard Arabic, it takes their role as the official language in almost all Arabic administrations. The effective spoken languages in daily conversations are Arabic dialects, which are spoken languages without a standardized writing form [2]. They can be classified into: Levantine (spoken in Palestine, Jordan, Syrian and Lebanon), Egyptian (in Egypt and Sudan), Maghrebi (spoken in the Arab Maghreb) and Iraqi [22], this later one may be also divided into Iraqi versus Gulf classes [37].

In these Dialect families, we will find also sub-families. In the case of the Algerian dialect, the work of [19] classify Algerian dialects into four groups:

1. the dialect of Algiers and its outskirts.
2. the dialect of the east in Annaba and its outskirts.
3. the dialect of Oran and the west of Algeria.
4. the dialect of the Algerian Sahara.

Even Algerian newspapers content is written in MSA and comments follow generally this style, we find some visitors that use Algerian Dialects words in their comments. For example the Arabic sentence: أشياء كهذه تحدث فقط في الدول المتخلفة Âaš.yAÂ kahaðihi taH.duΘu faqat fi Ald~uwal almutaxal~ifa[1] (things like this occur only in retarded countries) is written in a comment اشياء كيما هاذي تصرى غير في الدول المتخلفة Aaš.yA kima hAðy tas.ra γir fy Ald~uwal almutaxal~ifa.

---

[1]For transliteration we are following in this work the scheme developed by [16].

Also we found in several cases the use of the letter د d, instead of ذ ð, which is a characteristic of the Dialect of Algiers the capital of Algeria [19], as the case in the comment شكرا يا حفيظ هدا هو حال المسؤل الدى اسندة له المهمة و فشل اصبح ينتقد من اجل ان ينتقد و يطبل و يدافع عن الزمن الدى مر ولكن هناك رجال يصنعون المجد بتحديهم الوقائع šukrã ya HafiyĎ hada huwa HaAl Almasŵul Al~adi Âus.nidaħ. lahu Almuhim~aħ wa fašil. Asbaha yantaqid min Âjl Ân yantaqid wa yuTab~il wa yudaAfiς ςan Alz~aman Al~adi mar~ wa lakin hunAka rijAl yaSnaςuwn Almajd bitaHad~iyhim AlwaqaAĂς (Thank you hafid this is the state of the responsible to whom is affected a mission and he fails, so he become critic for critic and he defend the past time but there are men making the glory by confronting the realities).

In this paper, we are interested by comments in the Arabic Algerian online press, in the goal of developing an approach to classify these comments into positive and negative classes.

The paper is organized as follows. In the section 2 a literature review is given. In the section 3 a background of adopted methodology and used parameters are given. Section 4 is dedicated to the proposed approach. An experimental study is explained and obtained results are discussed in the section 5. We finish by conclusion and perspectives to future works.

## 2. Literature Review

Sentiment Analysis is an emergent field in the domains of Data Mining and Natural Language Processing (NLP); it is a research issue with the purpose of extract meaningful knowledge from user-generated content, for tracking the mood of people about events, products or topics [35]. It may be tacked as a classification problem, where the goal is to determine whether a written document, e.g. reviews, expresses a positive or negative opinion about specific entities [5, 24]. Works in sentiment analysis domain can be classified into statistical approaches also known as lexical based and supervised learning or corpus-based approaches [7]. In lexical approaches, we need sentiment lexical as resources to predict document sentiments, while in the supervised learning approaches, which will be used in this work, it is necessary to have well annotated corpora to create classification models.

For resources creation, we found several works. In [3] and [4] the authors build manually a Standard Arabic corpus SA from maktoob yahoo!, a corpus of 1.442 Arabic review covering five domains, with 50 topics in each domain. Reviews are written in different Arabic dialects in addition to MSA. The annotation is performed by three Arabic native speakers with different skills in computer science [4], and a consensus of 51,60% where the three annotators agree in one label is obtained. The best Inter Annotators Agreement (IAA) in term of kappa coefficient is achieved between annotators A2 and A3 as k= 0.556 considered as moderate. The works of [1] and [20] present a multi genre corpus for Modern Standard Arabic annotated at the sentence level. In [1], several annotation methods were adopted, and kappa parameter is used to measure IAA. The authors conclude that a training of annotators is necessary to have a consistent annotation. In [20], three Arabic native speakers annotators specialized in Arabic linguistics were working to annotate their corpus into positive, negative and neutral sentiment. An agreement of 97% is obtained over the annotators. Some rules are given to annotators to achieve a high degree of consistency. In the goal of creating an annotated corpus and lexicon, the authors in [18] have collected 15.274 Arabic reviews concern 13 governmental services. For annotation, thirty Arabic speakers with different skills and from different Arab countries are asked to annotate the reviews at two levels, i.e. linguistic issues and dialects. A high Inter Annotator Agreement of 0,78 in Kappa score is abstained. To decide the final annotation, three experts work to give the decision. In [30], the authors have created ARAACOM, ARAbic Algerian Corpus for Opinion Mining, 92 comments were collected from an Algerian Arabic newspaper web site. Support vector machines and naïve Bayes classifiers were used. Both unigram and bigram word model were tested. Best results are obtained in term of precision, and the bigram model increase results in almost cases. BRAD (Book Review in Arabic Dataset) [14], a corpus of 510.598 reviews expressed in MSA and Arabic dialects is collected from 'www.goodreads.com'. For annotation, the review with 1 or 2 stars is considered negative and with 4 or 5 stars is considered as positive. Neutral reviews with 3 stars are eliminated from the corpus. Four classifiers were used to test the performance of the created corpus. The best results are obtained with Support-Vector Machines (SVM) and Logistic Regression. In [12], 625 Arabic reviews and comments on hotels are collected from Trip Advisor website and classified into five categories: "ممتاز" (excellent); "جيد جدا" (very good); "متوسط" (middling); "ضعيف" (weak) and "مروع" (horrible). The modeling approach combining SVM with K-Nearest Neighbours (KNN) provides the best result (F-measure of 97%). In [15], HARD (Hotel Arabic-Reviews Dataset) a corpus composed of 490.587 reviews has been created. HARD covers 1.858 hotels and commented by 30.889 users. The authors come to a balanced dataset of 94.052 reviews form which 46.968 are annotated as positive and 47.084 as negative. The annotation is done following the rating system where reviews with rating 1 or 2 are considered negative and those of 4 or 5 rating are considered as positive, neutral comments with 3 rating are not considered in the dataset. For classification, six commonly used methods are applied, in addition to the use of a lexicon constructed in the same work achieving an accuracy of 89%. Logistic regression and SVM give the best results with a range of polarity classification from 94% to 97%. The authors of Curras [22] investigate in a corpus creation for Palestinian Arabic dialect. Two annotators are solicited to annotate morphologically Curras at the word level and Inter Annotators Agreement is calculated using Kappa coefficient. After annotation, both annotators work together to agree in the resultant gold standard. The best accuracy among the annotators achieves 98,8%. In [32], the authors create OCA (Opinion Corpus for Arabic) with 250 positive documents and 250 negative ones. The corpus is annotated at the document level by using web sites rating systems. Support vector machines and naïve Bayes classifiers were used for evaluation. The corpus documents are mostly related to movie reviews. The work [16] is based on the creation of a lexical resource to be used in the automatic annotation of a sentiment corpus for Algeria Dialect (AD) with positive and negative tags. The corpus is formed by Arabic and Arabizi scripts. Several machine learning methods are used to test the performance of the corpus and the Arabic text results are better than Arabizi text.

From this review of literature in opinion mining works and especially works dealing with Arabic language we can conclude that an important part of work concern specific topics. So, conducting studies with other topics require developing dedicated benchmarks that can be used to validate or revise existing results. Also, publicly available corpora are very sparse which makes very necessary the development of dedicated resources to carry out studies is this language.

# 3. Background

## 3.1. MATTER Approach

MATTER is a cyclic approach for annotate natural language texts, the approach is based on several iterations to achieve the annotation process [29]. The approach MATTER consists on a cycle of six steps. The model of the phenomenon may be revised for further train and test steps [21]:

1. Model: the first step is about modelling the studied phenomenon.
2. Annotate: an annotation can be seen as a meta data [28]. This metadata will be added to the corpus for data classification into predefined classes like positive, negative, neutral, etc., The annotation can be done at several levels. At document level [32], at sentence level [10] or at word level, also known as Part Of Speech tagging (POS) [22, 34]. We can find several ways to achieve annotation with; annotation by 2 to 5 persons having some specified skills [5, 29], Crowdsourcing where the annotation is done by an important number of annotators without specific skills [9] or annotation based on rating systems offered by opinion sites [32]. The final version of the annotated data, called the gold standard, is the corpus to be used in the classification step [29].
3. Train: a classifier would be trained with a part of the data.
4. Test: the rest of data (which is not used for training) is submitted to classifier for test.
5. Evaluate: evaluation metrics are calculated, to measure the annotation and classification performances.
6. Revise: based on evaluation metrics the model may be revised, and additional iteration is to do if needed.

## 3.2. Validation Method

In the scope of this work, the 10-fold cross-validation method is to be used. Cross-validation is a statistical method of evaluating and comparing learning algorithms by dividing the data set into two segments: one used to learn or train a model and the other used to validate the model. The basic form of cross-validation is k-fold cross-validation, in which the data set is first divided into k equally sized segments or folds. Subsequently, k iterations of training and validation are performed, so that, at each iteration a different fold of the data is held out for validation, while the remaining k-1 folds are used for learning [31]. The performance values are taken as a combination of the k performance values (as an average or another combination) to have a single estimation [27]. The authors in [23] and [33] conclude that 10-fold cross validation is the best alternative to follow in classification process, even if computation power allows more folds.

## 3.3. Classifiers

In this paper, three well known classifiers has been used. Those are:

1. Support-Vector Machines: SVM is a relatively new machine learning method for binary classification problems [13]. To have the best results with SVM the practitioner needs to well choice and fixe certain parameters: used kernel, gamma, and also well data collecting and pre-processing [8].
2. Naive Bayes: NB is based on the "Bayes assumption" in which the document is assigned to the class in which it belongs with the highest probability [26].
3. K-Nearest Neighbours: KNN is a simple classifier that uses an historical value search to find the future ones [36].

## 3.4. Evaluation Measures

1. Inter Annotators Agreement: several metrics are used in literature to evaluate IAA. The kappa coefficient [11] is the most used in two annotators based works[5, 29]. The coefficient is defined as :

$$k = \frac{\Pr(a) - \Pr(e)}{1 - \Pr(e)} \quad (1)$$

Where, Pr(a) represents the proportion of the cases in which both annotators agree, and Pr(e) is the proportion we search that the two annotators agree by chance [11].

2. Precision and Recall:

$$Precision = \frac{\text{TP}}{\text{TP+FP}} \quad (2)$$

$$Recall = \frac{\text{TP}}{\text{TP+FN}} \quad (3)$$

3. Accuracy: precision and recall are both complementary one to the other, we combine the two using the Accuracy measure given as:

$$Accuracy = \frac{\text{TP+TN}}{\text{TP+FP+TN+FN}} \quad (4)$$

- TP count the comments correctly assigned to the positive category.
- FP count the comments incorrectly assigned to the positive category.
- FN comments incorrectly rejected from the positive category.
- TN comments correctly rejected from the positive category.

## 4. Proposed Approach

Our proposed approach constitutes an enhancement of an existing approach ARAACOM (ARAbic Algerian Corous for Opinion Mining) [30].

In this work, MATTER approach [29] for comments annotation is enhanced. We add a processing (process) step to have MApTTER (Model Annotate process Train Test Evaluate Revise) approach. This enhancement allows us to give comments in the brut form to our annotators. So the processing step is included to the approach to:

- The annotators deal with the original text.
- The new examples can be added to any iteration.

For corpus creation, we collect comments from three Algerian Arabic newspapers web sites, in occurrence Echorouk[2], Elkhabar[3], and Ennahar[4]. We selected articles covering several subjects (news, political, religion, sports, and society).

The following algorithm summarizes our proposed approach:

*Algorithm 1: Our proposed approach*

```
Algorithm: Proposed approach
Begin
   IAA =0;
  while (IAA <= 100% ) do
       read (URL);
       Page =load (URL);
          while (there are comments in Page) do
               Extract the following Comment
             if (Comment in Data_base) then
                  Delete Comment ;
             Else
                  Add Comment to the Data_base;
             end if
          end while
       MODEL the phenomenon
       ANNOTATE
       Calculate New_IAA
          if New_IAA <= IAA then
               go to MODEL
          end if
       PROCESS
       TRAIN And TEST
       EVALUATE
          if (results sufficient)
             Break;
          Else
               REVISE
          end if
   end while
end
```

### 4.1. Models

The model is defined as the triplet: M= {T, R, I}

T= {Comment-classe, Positive, Negative, Neutral}: a set of terms.

R={Comment-classe::= Positive| Negative| Neutral}: a set of relations between terms.

I= {Positive: "Subjective with positive sentiment",
Negative: "Subjective with negative sentiment",
Neutral: "out of topic or without sentiment (objective)"}: interpretation of terms.

### 4.2. Annotate

Two Arabic native speakers are requested to annotate our corpus. In the beginning of each annotation round, a set of guidelines were given to annotators to have the best degree of contingency in obtained results.

1. Annotation Guidelines: In the guidelines the project must be described with its methodology, outcomes and all information needed to achieve our goals [21]. In each round of the MApTTER cycle, annotation guidelines will be refined taking into account previous results.
2. Adjudication: In adjudication, the annotations from different annotators are merged to have a single corpus called gold standard [21].

---

[2] www.echoroukonline.com/ara/
[3] www.elkhabar.com
[4] www.ennaharonline.com

### 4.3. Process

To have the best results in stemming and optimizing the word vector, a set of preprocessing steps are to conduct:

1. Manual text pre-processing: A lot of spelling mistakes are found, also some comments are written in languages other than MSA, such us French and English. First, all comments are translated into MSA, as samples the comment written in French language: 'merci Mr Hafid vous avez bien résumé qu est ce qui ce passe dans notre football' (Thank you hafid you resume well what happens in our football) is translated as شكرا حفيظ لقد لخصت جيدا ما يحدث في كرة القدم عندنا šukrã HafiyĎ laqad lax~aSta jay~idã ma yaHduθu fi kurah Alqadam ɕindanA, and the comment 'thank you' written in English is translated as شكرا لك šukrã lak.

Second, Repeated letters such as اليومممممممممممممممممممممممممم Alyaw.mmmmmmmmmm (today with the last letter repeated) becomes اليوم Alyaw.m (today). and بعييييييييييييييييييييييييدا baçiiiiiiiiiiiiiiiiiiiiiiiiydã (far with a middle letter repeated) which becomes بعيدا baçiydã (far).

Then Arabizi comments, when Arabizi is an Arabic language used in SMS and in tchat in Internet. It differs from transliteration that there is no standard to adopt in this language. Arabizi comments, in this work, are transformed into their Arabic equivalent. For example, the comment 'YA3TIK SAHA KHOYA BARATLI KHATRI FI L3ADYAN VIVE L'ALGÉRIE 1.2.3' (Thank you brother you Warmed my heart in the enemies, life to Algeria) is transformed to يعطيك الصحة أخي أثلجت خاطري في الأعداء، تحيا الجزائر yaçTik AlSiHah Âaxi Âθ.lajta xATiriy fi AlÂç.dA' taHyA AljazAŷir.

We finish by character encoding where all texts are resolved to UTF-8 encoding format.

2. Tokenization: in tokenization, words are separated by non-letters characters.
3. Stemming: light stemming is used in this step. Figure 1 and Figure 2 shows light stemming and stemming of the same comment. We remark for example that the word الجزائر AljazAŷir (Algeria with the definite article) in light stemming is stemmed to جزائر jazAŷir (Algeria without the definite article), when in stemming is stemmed as جزر jazar or jaz.r (carrot or ebb). And the word اقتصاديا Ăq.tiSAdiyã (economically) in light stemming is stemmed as it without changing اقتصاديا Ăq.tiSAdiyã, when in stemming it is stemmed as قصد qaSada (intention or meaning). In the scope of this work, we have used light stemming, because stemming generates the root of the word which gives a different meaning in several cases [27].

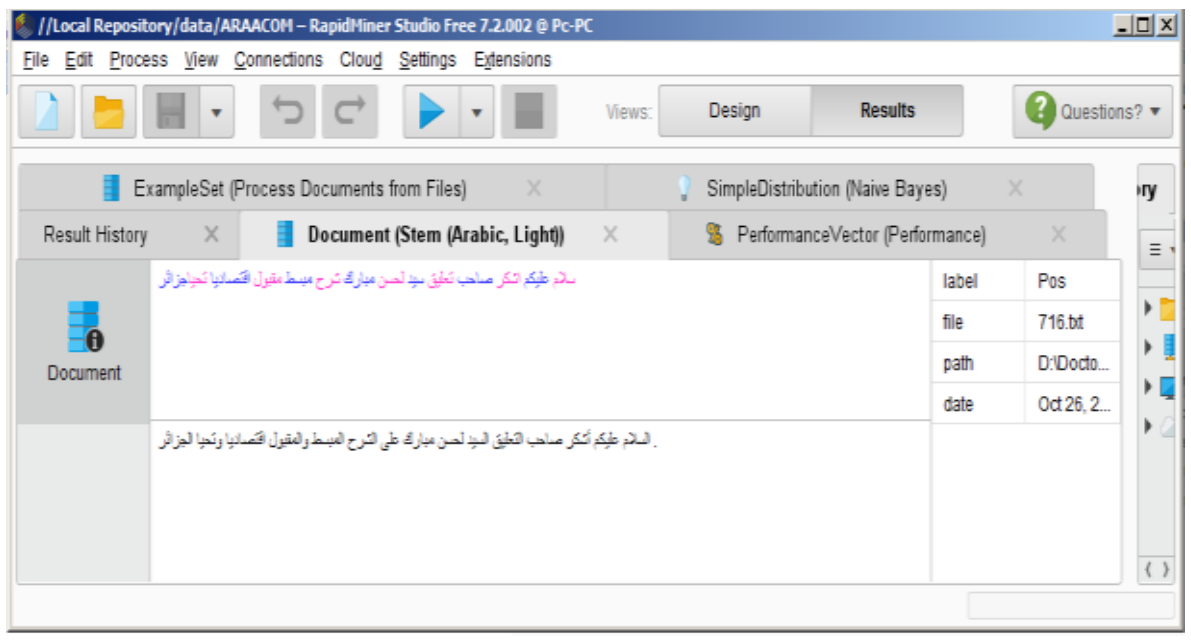

Figure 1. Sample of word light stemming.

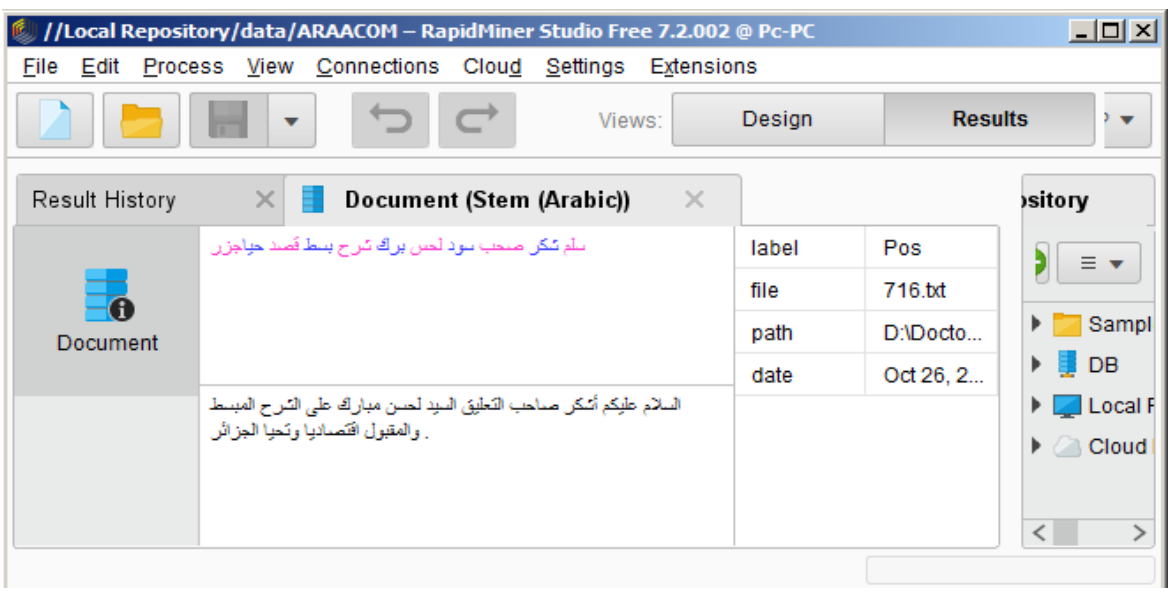

Figure 2. Sample of word Stemming.

4. Stop words Removal: a list of stop words is offered by the used toolkit.
5. Word n-gram: Unigram, bigram and trigram word are generated.
6. Word vector: four-word vectors are tested Term Frequency (TF), Term Occurrence (TO), Term Frequency Inverse Document Frequency (TF-IDF) and Binary Term Occurrence (BTO).

### 4.4. Train and Test

In this step three classification methods, SVM, NB, and KNN were used. And the 10-fold cross validation is adopted. For the SVM, we use SVM linear, and for KNN we adopt the k=9 parameter as suggested in [10].

### 4.5. Evaluate

For evaluation, Precision Recall and Accuracy are calculated for each classifier and obtained resultants were compared and discussed in each round of the MApTTER cycle.

### 4.6. Revise

In revision step adopted approach will be revised in the light of evaluation metrics, and a decision to continue or stop the MApTTER cycle is to take.

## 5. Experimental Study

### 5.1. First Round

1. Model: The above model is conserved as it. M={T,R,I}

T= {Comment-classe, Positive, Negative, Neutral}
R= {Comment-classe:: = Positive| Negative| Neutral}

I= {Positive: “Subjective with positive sentiment”, Negative: “Subjective with negative sentiment”, Neutral: “out of topic or without sentiment (objective)”}.

2. Annotate: The Annotation guidelines given to annotators are, to annotate a comment as it presents a positive, negative or neutral sentiment regarding the article topic, so for each comment the annotators have the correspondent article.
   - IAA Inter Annotators Agreement:
   We obtain : $k = 0.5195$ considered as moderate [29].
   - Adjudication: in adjudication step the two annotators are working together in the goal of obtaining a consensus in annotation, in cases when a consensus was not obtained the comment is considered as neutral. In this step, 45 comments are considered as positive, 88 as negative and 45 as neutral. We take in this work only the positive and negative classes, to have equilibrium we generate the corpus with the 45 positive comments and 45 of the negative ones.
3. Processing: The processing steps describing above were doing again, namely: text pre-processing, UTF-8 encoding, tokenization, stemming, stop words removal, n-gram word generation, and word vectors creation.
4. Train and Test: For these two steps, cross validation method was adopted using the three classifiers, support vector machines, naïve Bayes and k-nearest neighbours.
5. Evaluate: To evaluate our model we calculate precision and recall of both negative and positive classes and accuracy of the classification at whole.

Table1. First round Classification with SVM.

| Light Stem | | Unigram | | | | | Bigram | | | | | Tri-gram | | | | |
|---|---|---|---|---|---|---|---|---|---|---|---|---|---|---|---|---|
| | | Acc | Precision | | Recall | | Acc | Precision | | Recall | | Acc | Precision | | Recall | |
| | | | Pos | Neg | pos | Neg | | pos | Neg | pos | Neg | | pos | Neg | pos | Neg |
| No | TO | 58,89 | 55,71 | 70,00 | 86,67 | 31,11 | 58,89 | 55,71 | 70,00 | 86,67 | 31,11 | 57,78 | 55,07 | 66,67 | 84,44 | 31,11 |
| | TF | 55,56 | 53,16 | 72,73 | 93,33 | 17,78 | 54,44 | 52,44 | 75,00 | 95,56 | 13,33 | 54,44 | 52,44 | 75,00 | 95,56 | 13,33 |
| | TF-IDF | 52,22 | 51,22 | 62,50 | 93,33 | 11,11 | 52,22 | 51,22 | 62,50 | 93,33 | 11,11 | 53,33 | 51,85 | 66,67 | 93,33 | 13,33 |
| | BTO | 60,00 | 56,52 | 71,43 | 86,67 | 33,33 | 58,89 | 55,71 | 70,00 | 86,67 | 31,11 | 57,78 | 54,93 | 68,42 | 86,67 | 28,89 |
| Yes | TO | 63,33 | 59,38 | 73,08 | 84,44 | 42,22 | 58,89 | 55,71 | 70,00 | 86,67 | 31,11 | 60,00 | 56,34 | 73,68 | 88,89 | 31,11 |
| | TF | 70,00 | 66,07 | 76,47 | 82,22 | 57,78 | 65,56 | 61,67 | 73,33 | 82,22 | 48,89 | 66,67 | 62,71 | 74,19 | 82,22 | 51,11 |
| | TF-IDF | 67,78 | 64,29 | 73,53 | 80,00 | 55,56 | 67,78 | 64,81 | 72,22 | 77,78 | 57,78 | 68,89 | 65,45 | 74,29 | 80,00 | 57,78 |
| | BTO | 62,22 | 58,73 | 70,37 | 82,22 | 42,22 | 62,22 | 58,21 | 73,91 | 86,67 | 37,78 | 63,33 | 59,09 | 75,00 | 86,67 | 40,00 |

In Table1 are presented evaluations of SVM classifier. We observe that the stemming approach gives low performance comparing with the basic model and we report this result to the Algerian dialect. For example, we observed in the used tool that the name حفيظ HafiyĎ (which is a proper name widely used in Algeria and also mean in Arabic ‘conservator’) when it is stemmed take the root حفظ HafaĎa (to conserve). So stemming is not the best issue in such situations. Also, trigram does not give considerable amelioration in the most of results.

With the TF vector we obtain the best result as recall of the positive class (95%).

Table 2 shows classification results using Naïve Bayes method.

Table 2. First round Classification using NB.

| Light Stem | | Unigram | | | | | Bigram | | | | | Tri-gram | | | | |
|---|---|---|---|---|---|---|---|---|---|---|---|---|---|---|---|---|
| | | Acc | Precision | | Recall | | Acc | Precision | | Recall | | Acc | Precision | | Recall | |
| | | | Pos | Neg | pos | Neg | | Pos | Neg | pos | Neg | | Pos | Neg | pos | Neg |
| No | TO | 64,44 | 64,44 | 64,44 | 64,44 | 64,44 | 61,11 | 60,87 | 61,36 | 62,22 | 60,00 | 62,22 | 61,70 | 62,79 | 64,44 | 60,00 |
| | TF | 63,33 | 64,29 | 62,50 | 60,00 | 66,67 | 58,89 | 59,52 | 58,33 | 55,56 | 62,22 | 60,00 | 60,47 | 59,57 | 57,78 | 62,22 |
| | TF-IDF | 65,56 | 65,91 | 65,22 | 64,44 | 66,67 | 63,33 | 63,64 | 63,04 | 62,22 | 64,44 | 63,33 | 63,64 | 63,04 | 62,22 | 64,44 |
| | BTO | 58,89 | 58,70 | 59,09 | 60,00 | 57,78 | 57,78 | 57,45 | 58,14 | 60,00 | 55,56 | 58,89 | 58,33 | 59,52 | 62,22 | 55,56 |
| Yes | TO | 68,89 | 69,77 | 68,09 | 66,67 | 71,11 | 63,33 | 63,64 | 63,04 | 62,22 | 64,44 | 64,44 | 65,12 | 63,83 | 62,22 | 66,67 |
| | TF | 66,67 | 68,29 | 65,31 | 62,22 | 71,11 | 63,33 | 64,29 | 62,50 | 60,00 | 66,67 | 63,33 | 64,29 | 62,50 | 60,00 | 66,67 |
| | TF-IDF | 70.00 | 70,45 | 69,57 | 68,89 | 71,11 | 65,56 | 67,50 | 64,00 | 60,00 | 71,11 | 66,67 | 69,23 | 64,71 | 60,00 | 73,33 |
| | BTO | 66,67 | 66,67 | 66,67 | 66,67 | 66,67 | 61,11 | 60,87 | 61,36 | 62,22 | 60,00 | 62,22 | 62,22 | 62,22 | 62,22 | 62,22 |

Comparing to SVM, obtained results with Naive Bayes are low and are not competitive.

Table 3. First round Classification using KNN.

| Light Stem | | Unigram | | | | | Bigram | | | | | Tri-gram | | | | |
|---|---|---|---|---|---|---|---|---|---|---|---|---|---|---|---|---|
| | | Acc | Precision | | Recall | | Acc | Precision | | Recall | | Acc | Precision | | Recall | |
| | | | Pos | Neg | pos | Neg | | Pos | Neg | pos | Neg | | Pos | Neg | pos | Neg |
| No | TO | 52,22 | 51,16 | 75,00 | 97,78 | 6,67 | 51,11 | 50,57 | 66,67 | 97,78 | 4,44 | 51,11 | 50,57 | 66,67 | 97,78 | 4,44 |
| | TF | 57,78 | 55,74 | 62,07 | 75,56 | 40,00 | 56,67 | 55,17 | 59,38 | 71,11 | 42,22 | 56,67 | 55,17 | 59,38 | 71,11 | 42,22 |
| | TF-IDF | 63,33 | 60,71 | 67,65 | 75,56 | 51,11 | 63,33 | 60,71 | 67,65 | 75,56 | 51,11 | 61,11 | 58,93 | 64,71 | 73,33 | 48,89 |
| | BTO | 50,00 | 50,00 | 50,00 | 97,78 | 2,22 | 52,22 | 51,16 | 75,00 | 97,78 | 6,67 | 52,22 | 51,16 | 75,00 | 97,78 | 6,67 |
| Yes | TO | 48,89 | 49,32 | 47,06 | 80,00 | 17,78 | 51,11 | 50,63 | 54,55 | 88,89 | 13,33 | 52,22 | 51,25 | 60,00 | 91,11 | 13,33 |
| | TF | 70,00 | 68,00 | 72,50 | 75,56 | 64,44 | 68,89 | 67,35 | 70,73 | 73,33 | 64,44 | 68,89 | 67,35 | 70,73 | 73,33 | 64,44 |
| | TF-IDF | 64,44 | 65,12 | 63,83 | 62,22 | 66,67 | 64,44 | 65,85 | 63,27 | 60,00 | 68,89 | 64,44 | 65,85 | 63,27 | 60,00 | 68,89 |
| | BTO | 53,33 | 51,72 | 100 | 100 | 6,67 | 51,11 | 50,56 | 100 | 100 | 2,22 | 51,11 | 50,56 | 100 | 100 | 2,22 |

K-nearest neighbours, as presented in Table 3. gives the best performance as classifier. And the best results are obtained in the recall of the positive class. The stemming gives some amelioration, but it is still not very important.

6. Revise: After evaluation of the IAA and classification scores, some suggestion are mentioned to take into account in the next round:
   - In the following annotation guidelines, the authors must take into account only the content of the comment no matter what the article subject is. So this recommendation may conduct to more consensuses between annotators, that in the previous round, when taking into account the article topic, we observe that for the same, comment one annotator can have a positive sentiment and the other have a neutral or even negative sentiment.

## 5.2. Second Round

1. Model: We conserve the above model without any changes.
2. Annotate: In this round we give our annotators instruction to consider only the content of the comment, as it is recommended in the revision of the above step.
   - IAA Inter Annotators Agreement: The Inter Annotator Agreement is improved with the new instruction, so it achieve k = 0.5820.
3. Adjudication*:* We generate our gold standard by adjudication of annotation step. And this is performed to obtain a single annotated corpus from the two available. We eliminate the neutral class and for equilibrium we take the 194 negative comments and only 194 from positives.
4. Processing: The above described processing steps are conducted twice.
5. Train and Test: We use always the three classifiers, SVM, NB and KNN. And the 10-fold cross-validation method.
6. Evaluate: From Table 4 Table in this second round we remark that even the IAA is improved, the SVM classification results are worse that the first round.

Table 4. Second round with SVM.

| Light Stem | | Unigram | | | | | Bigram | | | | | Tri-gram | | | | |
|---|---|---|---|---|---|---|---|---|---|---|---|---|---|---|---|---|
| | | Acc | Precision | | Recall | | Acc | Precision | | Recall | | Acc | Precision | | Recall | |
| | | | Pos | Neg | Pos | Neg | | Pos | Neg | Pos | Neg | | Pos | Neg | Pos | Neg |
| No | TO | 71,13 | 81,06 | 66,02 | 55,15 | 87,11 | 69,85 | 77,30 | 65,59 | 56,19 | 83,51 | 71,13 | 78,87 | 66,67 | 57,73 | 84,54 |
| | TF | 70,62 | 83,33 | 64,93 | 51,55 | 89,69 | 70,88 | 81,89 | 65,52 | 53,61 | 88,14 | 70,62 | 82,79 | 65,04 | 52,06 | 89,18 |
| | TF-IDF | 69,33 | 82,05 | 63,84 | 49,48 | 89,18 | 69,07 | 80,33 | 63,91 | 50,52 | 87,63 | 70,88 | 82,40 | 65,40 | 53,09 | 88,66 |
| | BTO | 71,39 | 81,68 | 66,15 | 55,15 | 87,63 | 70,10 | 79,55 | 65,23 | 54,12 | 86,08 | 68,56 | 76,87 | 64,17 | 53,09 | 84,02 |
| Yes | TO | 71,13 | 77,33 | 67,23 | 59,79 | 82,47 | 69,33 | 74,51 | 65,96 | 58,76 | 79,90 | 71,91 | 77,78 | 68,09 | 61,34 | 82,47 |
| | TF | 70,88 | 75,47 | 67,69 | 61,86 | 79,90 | 72,16 | 79,05 | 67,92 | 60,31 | 84,02 | 71,39 | 78,62 | 67,08 | 58,76 | 84,02 |
| | TF-IDF | 69,33 | 72,73 | 66,82 | 61,86 | 76,80 | 72,16 | 79,45 | 67,77 | 59,79 | 84,54 | 69,85 | 73,91 | 66,96 | 61,34 | 78,35 |
| | BTO | 71,65 | 77,27 | 67,95 | 61,34 | 81,96 | 69,33 | 74,19 | 66,09 | 59,28 | 79,38 | 72,16 | 77,22 | 68,70 | 62,89 | 81,44 |

Table 5. Second round classification using Naïve Bayes.

| Light Stem | | Unigram | | | | | Bigram | | | | | Tri-gram | | | | |
|---|---|---|---|---|---|---|---|---|---|---|---|---|---|---|---|---|
| | | Acc | Precision | | Recall | | Acc | Precision | | Recall | | Acc | Precision | | Recall | |
| | | | Pos | Neg | Pos | Neg | | Pos | Neg | Pos | Neg | | Pos | Neg | Pos | Neg |
| No | TO | 70,36 | 79,26 | 65,61 | 55,15 | 85,57 | 70,62 | 84,48 | 64,71 | 50,52 | 90,72 | 70,36 | 84,35 | 64,47 | 50,00 | 90,72 |
| | TF | 70,36 | 73,94 | 67,71 | 62,89 | 77,84 | 71,13 | 78,08 | 66,94 | 58,76 | 83,51 | 71,13 | 80,60 | 66,14 | 55,67 | 86,60 |
| | TF-IDF | 70,36 | 74,84 | 67,25 | 61,34 | 79,38 | 71,91 | 79,72 | 67,35 | 58,76 | 85,05 | 72,16 | 81,16 | 67,20 | 57,73 | 86,60 |
| | BTO | 72,16 | 83,59 | 66,54 | 55,15 | 89,18 | 71,65 | 88,18 | 65,11 | 50,00 | 93,30 | 70,88 | 87,85 | 64,41 | 48,45 | 93,30 |
| Yes | TO | 73,45 | 78,26 | 70,04 | 64,95 | 81,96 | 72,94 | 83,97 | 67,32 | 56,70 | 89,18 | 72,42 | 84,25 | 66,67 | 55,15 | 89,69 |
| | TF | 73,97 | 77,51 | 71,23 | 67,53 | 80,41 | 75,00 | 80,12 | 71,37 | 66,49 | 83,51 | 74,74 | 80,77 | 70,69 | 64,95 | 84,54 |
| | TF-IDF | 73,45 | 76,30 | 71,16 | 68,04 | 78,87 | 73,97 | 78,88 | 70,48 | 65,46 | 82,47 | 73,97 | 79,25 | 70,31 | 64,95 | 82,99 |
| | BTO | 74,74 | 83,33 | 69,67 | 61,86 | 87,63 | 74,23 | 87,30 | 67,94 | 56,70 | 91,75 | 73,71 | 87,10 | 67,42 | 55,67 | 91,75 |

In the case of the Naïve Bayes the obtained results are very promising; we achieve 93.30% of recall in the negative class without stemming, as shown in Table 5.

Table 6. Second round classification using KNN.

| Light Stem | | Unigram | | | | | Bigram | | | | | Tri-gram | | | | |
|---|---|---|---|---|---|---|---|---|---|---|---|---|---|---|---|---|
| | | Acc | Precision | | Recall | | Acc | Precision | | Recall | | Acc | Precision | | Recall | |
| | | | Pos | Neg | Pos | Neg | | Pos | Neg | Pos | Neg | | Pos | Neg | Pos | Neg |
| No | **TO** | 64,69 | 83,53 | 59,41 | 36,60 | **92,78** | 63,92 | 82,14 | 58,88 | 35,57 | **92,27** | 63,66 | 81,93 | 58,69 | 35,05 | **92,27** |
| | **TF** | 61,60 | 90,91 | 56,76 | 25,77 | **97,42** | 61,86 | 89,66 | 56,97 | 26,80 | **96,91** | 62,11 | 89,83 | 57,14 | 27,32 | **96,91** |
| | **TF-IDF** | 56,19 | 92,86 | 53,33 | 13,40 | **98,97** | 56,96 | 90,91 | 53,80 | 15,46 | **98,45** | 57,73 | 91,67 | 54,26 | 17,01 | **98,45** |
| | **BTO** | 63,92 | 92,19 | 58,33 | 30,41 | **97,42** | 62,89 | 89,06 | 57,72 | 29,38 | **96,39** | 62,37 | 88,71 | 57,36 | 28,35 | **96,39** |
| Yes | **TO** | 64,69 | 78,22 | 59,93 | 40,72 | 88,66 | 65,72 | 79,61 | 60,70 | 42,27 | 89,18 | 65,72 | 79,61 | 60,70 | 42,27 | 89,18 |
| | **TF** | **65,72** | 67,63 | 64,19 | 60,31 | 71,13 | **67,78** | 69,94 | 66,05 | 62,37 | 73,20 | **67,53** | 69,77 | 65,74 | 61,86 | 73,20 |
| | **TF-IDF** | 65,21 | 65,61 | 64,82 | 63,92 | 66,49 | 66,49 | 66,33 | 66,67 | 67,01 | 65,98 | 66,49 | 66,33 | 66,67 | 67,01 | 65,98 |
| | **BTO** | 63,66 | 85,33 | 58,47 | 32,99 | **94,33** | 62,63 | 81,01 | 57,93 | 32,99 | **92,27** | 62,63 | 81,01 | 57,93 | 32,99 | **92,27** |

In the case of the KNN classifier (see Table 6) the recall of the negative class is very good without stemming with all vectors.

7. Revise: In this point we decide to stop this cycle and report the obtained results.

## 6. Conclusions and Perspectives

Available corpora in the web for carried out Arabic sentiment analysis studies are very rare, and those available are generally related to movie reviews due to the available comments in correspondent websites. The case of newspapers comments is more delicate, that it is related to the country in which it is exist, so dealing with such comments must take into account, not only Arabic language but also related dialects and used languages in this country. In this work we proposed our enhanced approach for opinion mining in Algerian Arabic Newspapers comments.

For annotation, MApTTER approach is used to annotate Algerian newspaper comments. MApTTER is based on an existing annotation approach MATTER.

Three classifiers were used for a binary classification of comments into positive vs. negative, and the KNN one gives the most important performance. Obtained results are promising, but still to develop, and the most important conclusion is about stemming that do not improve the classification in our case. Also, the trigram model is not a good representation regarding the obtained results.

We aim in the future to develop the approach by considering the parts in the comment that enclose the most of semantic, which is the first and the last parts. Also taking into account non-Arabic parts of the comment may allow generalising the model.

The use of other techniques for sentiment analysis classification such as deep learning methods is a very important which we project to apply as continuation of this work.

## References

[1] Abdul-Mageed M. and Diab M., "AWATIF: A Multi-Genre Corpus for Modern Standard Arabic Subjectivity and Sentiment Analysis," *in Proceedings of Language Resources and Evaluation Conference*, Istanbul, pp. 3907-3914, 2012.

[2] Al-Harbi O., "Classifying Sentiment Of Dialectal Arabic Reviews: A Semi-Supervised Approach," *The International Arab Journal of Information Technology*, vol. 16, no. 6, pp. 995–1002, 2019.

[3] Al-Kabi M. N., Al-ayyoub M., Alsmadi I., and Wahsheh H., "A Prototype for a Standard Arabic Sentiment Analysis Corpus," *International Arab Journal of Information Technology*, vol. 13, no. 2, pp. 163-170, 2016.

[4] Al-Kabi M., Al-qwaqenah A., Gigieh A., Alsmearat K., Al-ayyoub M., and Alsmadi I., "Building a Standard Dataset for Arabic Sentiment Analysis," *in Proceedings of the IEEE/ACS 13th International Conference of Computer Systems and Applications*,Agadir, pp. 1-6, 2016.

[5] Alotaibi S. and Anderson C., "Extending the Knowledge of the Arabic Sentiment Classification Using a Foreign External Lexical Source," *International Journal on Natural Language Computing*, vol. 5, no. 3, pp. 1-11, 2016.

[6] Atia S. and Shaalan K., "Increasing the accuracy of opinion mining in Arabic," *in Proceedings of Conference: International Conference on Arabic Computational Linguistics*, Cairo, pp. 106-113, 2015.

[7] Badaro G., Baly R., Hajj H., El-Hajj W., Shaban K ., Habash N ., Al-Sallab A., and Hamdi A., "A Survey of Opinion Mining in Arabic: A Comprehensive System Perspective Covering Challenges and Advances in Tools, Resources, Models, Applications, and Visualizations," *ACM Transactions on Asian Language Information Processing*, vol. 18, no. 3, pp. 1-52, 2019.

[8] Ben-Hur A. and Weston J., "A user's guide to support vector machines.," *Methods in molecular biology*, vol. 609, pp. 223–239, 2010.

[9] Bougrine S., Cherroun H., and Abdelali A., "Altruistic Crowdsourcing for Arabic Speech

Corpus Annotation," *Procedia Computer Science*, vol. 117, pp. 133-144, 2017.

[10] Brahimi B., Touahria M., and Tari A., "Data and Text Mining Techniques for Classifying Arabic Tweet Polarity," *Journal of Digital Information Management*, vol. 14, no. 1, pp. 15-25, 2016.

[11] Carletta J., "Assessing agreement on classification tasks: the kappa statistic," Computational Linguistics, vol. 22, no. 2, pp. 249-254, 1996.

[12] Cherif W., Madani A., and Kissi M., "Towards an Efficient Opinion Measurement in Arabic Comments," *Procedia Comput Science*, vol. 73, pp. 122-129, 2015.

[13] Cortes C. and Vapnik V., "Support-Vector Networks," *Machine Learning*, vol. 20, no. 3, pp. 273-297, 1995.

[14] Elnagar A. and Einea O., "BRAD 1 . 0 : Book Reviews in Arabic Dataset," *in Proceedings of IEEE/ACS 13th International Conference of Computer Systems and Applications*, Agadir, pp. 1-8, 2016.

[15] Elnagar A., Khalifa Y., and Einea A., *Intelligent Natural Language Processing: Trends and Applications*, in Springer International Publishing, 2018.

[16] Guellil I., Adeel A., Azouaou F., and Hussain A., "SentiALG : Automated Corpus Annotation for Algerian Sentiment Analysis Introduction," *in Proceedings of the International Conference on Brain Inspired Cognitive Systems*, Xi'an, pp. 557-567, 2018

[17] Habash N., Soudi A., and Buckwalter T., "On Arabic Transliteration," *In Arabic Computational Morphology: Knowledge-based and Empirical Methods*, vol. 49, no. 4, pp. 15-22, 2007.

[18] Hamdi A., Shaban K., and Zainal A., "CLASENTI : A class-specific sentiment analysis framework," *ACM Transactions on Asian and Low-Resource Language Information Processing*, vol. 17, no. 4, p. 32, 2018.

[19] Harrat S., Meftouh K., Abbas M., Hidouci K., and Smaili K., "An Algerian Dialect : Study and Resources," *International Journal of Advanced Computer Science and Applications*, vol. 7, no. 3, pp. 384-396, 2016.

[20] Ibrahim H. S., Abdou S., and Gheith M., "MIKA: A Tagged Corpus for Modern Standard Arabic and Colloquial Sentiment Analysis," *in Proceedings of IEEE 2nd International Conference on Recent Trends in Information Systems,* Kolkata, pp 353-358, 2015.

[21] Ide N. and Pustejovsky J., *Handbook of Linguistic Annotation*, Springer, 2017.

[22] Jarrar M., Habash N., Alrimawi F., Akra D., and Zalmout N., "Curras: an annotated corpus for the Palestinian Arabic dialect," *Language Resources and Evaluation*, vol. 51, no. 3, pp. 745-775, 2017.

[23] Kohavi R., "A Study of Cross-Validation and Bootstrap for Accuracy Estimation and Model Selection," *in Proceedings of the 14th international joint conference on Artificial intelligence*, Montreal, pp. 1137-1143, 1995.

[24] Korayem M., Aljadda K., and D. Crandall, "Sentiment/subjectivity analysis survey for languages other than English," *Social Network Analysis and Mining*, vol. 6, no. 1, pp. 1-17, 2016.

[25] Liu B., "Sentiment Analysis and Opinion Mining," *in Proceedings of Synthesis Lectures on Human Language Technologies*, pp. 1-167, 2012.

[26] McCallum A. and Nigam K., "A Comparison of Event Models for Naive Bayes Text Classification," *in Proceedings of AAAI-98 workshop on learning for text categorization*, Madison, pp. 41-48, 1998.

[27] Mountassir A., Benbrahim H., and Berraba I., "Sentiment Classification on Arabic Corpora. A Preliminary Cross-Study," *Document Numerique*, vol. 16, no. 1, pp.73-96, 2013.

[28] Petrillo M. and Baycroft J., "Introduction to Manual Annotation," *Fairview Research*, pp 1-7, 2010,

[29] Pustejovsky J. and A. Stubbs, *Natural Language Annotation for Machine Learning*, O'Reilly Media, 2012.

[30] Rahab H., Zitouni A., and Djoudi M., "ARAACOM: ARAbic Algerian Corpus for Opinion Mining," *in Proceedings of the International Conference on Computing for Engineering and Sciences*, Istanbul, pp. 35-39, 2017.

[31] Refaeilzadeh P., Tang L., and Liu H., "Cross-Validation," *in Encyclopedia of Database Systems*, Boston, pp. 532–538, 2009.

[32] Rushdi-Saleh M., Martín-Valdivia M., Ureña López L., and Perea-Ortega J., "OCA: Opinion Corpus for Arabic," *Journal of the American Society for Information Science and Technology*, vol. 62, no. 10, pp. 2045-2054, 2011.

[33] Salzberg S., "On Comparing Classifiers : Pitfalls to Avoid and a Recommended Approach," *Data Mining and Knowledge Discovery*, vol. 1, pp. 317-328, 1997.

[34] Tunga G., "Part-of-Speech Tagging," in *Handbook of Natural Language Processing*, Second Edi., CRC Press, pp. 205–235, 2010.

[35] Vinodhini G. and Chandrasekaran R., "Sentiment Analysis and Opinion Mining : A Survey," *International Journal of Advanced Research in Computer Science and Software Engineering* , vol. 2, no. 6, pp. 282-292, 2012.

[36] Wang X., An K., Tang L., and Chen X.,"Short Term Prediction of Freeway Exiting Volume

Based on SVM and KNN," *International Journal of Transportation Science and Technology*, vol. 4, no. 3, pp. 337-352, 2015.

[37] Zaidan O. and Callison-burch C., "The Arabic Online Commentary Dataset : an Annotated Dataset of Informal Arabic with High Dialectal Content," *in Proceedings of the 49th Annual Meeting of the Association for Computational Linguistics*, Portland, pp. 37-41, 2011.

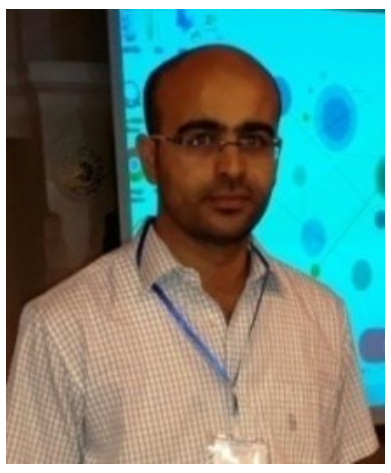

**Hichem Rahab** is currently working as an Assistant Professor in department of Mathematics and computer science in the University of Khenchela, Algeria. He obtained his master's degree in computer science from Batna University, Algeria, 2012. His research interest includes machine learning, Arabic opinion mining and sentiment analysis.

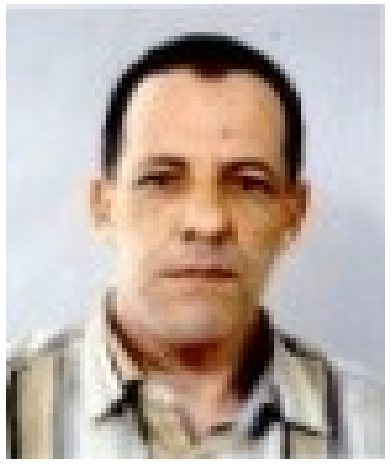

**Abdelhafid Zitouni** received his PhD in computer science in 2008 from the University of Constantine, Algeria. He is currently working as Professor in University of Constantine 2 Abdelhamid Mehri. His research interests include Cloud Computing, Security, and Arabic text mining field. Pr. Abdelhafid Zitouni has published many articles in International Journals and Conferences. He peer-reviewed conference and journal papers in the above research topics.

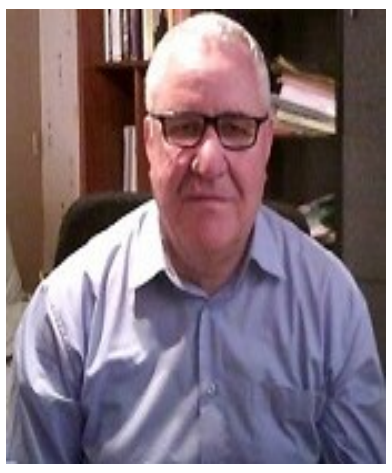

**Mahieddine Djoudi** received a PhD in Computer Science from the University of Nancy, France, in 1991. His PhD thesis research was in Acoustic Phonetic Decoding for Standard Arabic Speech Recognition. He is currently working at Computer Science Department, Faculty of Fundamental and Applied Sciences at the University of Poitiers, France and member of TechNE Technology Enhanced Learning Research Laboratory. His main scientific interests are e-Learning, Mobile Learning, Cloud Computing, Information Literacy and Learning Analytics. He has published over 100 scientific papers. He is also a member of program committees, editor or reviewer for international journals or conferences proceedings.